\documentclass[11pt]{article}
\usepackage{fullpage}
\usepackage{cite}
\usepackage{amsmath}
\usepackage{amssymb}
\usepackage{graphicx}
\usepackage[utf8]{inputenc} 

\title{}
\date{}
\def\para{\\ [-2mm]}

\def \be  {\begin{equation}}
\def \ee  {\end{equation}}
\def \ba  {\begin{eqnarray}}
\def \ea  {\end{eqnarray}}
\newcommand{\nn}{\nonumber}
\def\eqn#1{eq.~(\ref{#1})} \def\Eqn#1{Equation~(\ref{#1})}
\def\eqns#1#2{eqs.~(\ref{#1}) and~(\ref{#2})}

\def\IZ{\relax\ifmmode\mathchoice
{\hbox{\cmss Z\kern-.4em Z}}{\hbox{\cmss Z\kern-.4em Z}}
{\lower.4pt\hbox{\cmsss Z\kern-.4em Z}}
{\lower1.2pt\hbox{\cmsss Z\kern-.4em Z}}\else{\cmss Z\kern-.4em Z}\fi}
\newcommand{\Z}{\mathsf{Z}\kern -5pt \mathsf{Z}}
\newcommand{\unit}{\mathsf{1}\kern -3pt \mathsf{l}}

\def\eg{{e.g.}}

\def\s { s }
\def\barpsi {{\bar\psi}}
\def\X {{\barpsi}}
\def\Y {{\psi}}

\def\cA { {\cal A}  }

\def\Agv{ \cA^{\rm grav}_n }
\def\Agluon{ \cA^{\rm gluon}_n }
\def\Abiad{ \cA^{\rm biadjoint}_n }

\def\Ank  { \cA^{\rm qcd}_{n,k} }
\def\Ankl { \cA^{\rm qcd}_{n,k\le 2} }
\def\Abi{ \cA^{\rm bicolor} }
\def\Abink{ \cA^{\rm bicolor}_{n,k} }
\def\Abinkl{ \cA^{\rm bicolor}_{n,k\le 2} }
\def\Agvnk  { \cA^{\rm grav}_{n,k} }
\def\Agvnkl{ \cA^{\rm grav}_{n,k\le 2} }

\def\A {  A^{{\rm (s)}} }
\def\tA {  {\tilde  A}^{{\rm (s)}} }
\def\tAYM {  {\tilde  A} }

\def\CF { {C_{1\gamma 2}} }
\def\NF { {N_{1\gamma 2}} }
\def\NFp { {N'_{1\gamma 2}} }
\def\CFp { {C'_{1\gamma 2}} }
\def\CFps { {C'_{13\sigma  2}} }
\def\NFps { {N'_{13\sigma  2}} }
\def\tCF { {\tilde{C}_{1\delta 2}} }
 
\def\tc{\tilde{c}}

\def\ta {\textsf{a}}
\def\tb {\textsf{b}}
\def\ti {\textsf{i}}

\def\dela {\delta_{a}}
\def\tdela {{\tilde \delta}_{a}}

\def\MB{ {\rm Melia} }
\def\JO{ {\rm JO} }

\begin{document}

\titlepage
\begin{flushright}
BOW-PH-166\\
\end{flushright}

\vspace{3mm}

\begin{center}

{\Large\bf\sf
KLT-type relations for QCD 
and bicolor amplitudes
\\ [2mm]
from color-factor symmetry
}

\vskip 3cm

{\sc
Robert W. Brown$^a$
and Stephen G. Naculich$^{b}$
}

\vskip 0.5cm
$^a${\it
Department of Physics\\
Case Western Reserve University\\
Cleveland, OH 44106 USA
}

\vskip 0.5cm
$^b${\it
Department of Physics\\
Bowdoin College\\
Brunswick, ME 04011 USA
}

\vspace{5mm}
{\tt
rwb@case.edu,  naculich@bowdoin.edu
}
\end{center}

\vskip 3cm

\begin{abstract}

Color-factor symmetry is used to derive a 
KLT-type relation for tree-level QCD amplitudes
containing gluons and an arbitrary number of massive or
massless quark-antiquark pairs,
generalizing the expression for Yang-Mills amplitudes 
originally postulated by Bern, De Freitas, and Wong.
An explicit expression is given for all amplitudes 
with two or fewer quark-antiquark pairs
in terms of the (modified) momentum kernel.

We also introduce the bicolor scalar theory,
the ``zeroth copy'' of QCD, 
containing massless biadjoint scalars and 
massive  bifundamental scalars,
generalizing the biadjoint scalar theory of
Cachazo, He, and Yuan.
We derive KLT-type relations for tree-level
amplitudes of biadjoint and bicolor theories
using the color-factor symmetry possessed by these theories.

\end{abstract}

\vspace*{0.5cm}

\vfil\break

\section{Introduction}
\setcounter{equation}{0}

Over thirty years ago, Kawai, Lewellen, and Tye (KLT)
discovered that tree-level closed-string scattering amplitudes 
can be expressed as a sum of products of
open-string scattering amplitudes \cite{Kawai:1985xq}.
In the field-theory  limit, the KLT formula relates
gravitational scattering amplitudes to products
of gauge-theory scattering amplitudes \cite{Berends:1988zp,Bern:1998sv,BjerrumBohr:2010ta,BjerrumBohr:2010zb,BjerrumBohr:2010yc,BjerrumBohr:2010hn,Broedel:2013tta,Cachazo:2013iea}.
The tree-level $n$-graviton amplitude may be written in the compact form
\be
\Agv
~= ~-~ \sum_{\sigma, \tau \in S_{n-3}} 
A (1,3, \sigma, 2)  ~S[\sigma|\tau]_3  ~A (2,  3,   \tau, 1)
\label{KLT}
\ee
where $A(\cdots)$ denotes color-ordered (or partial) $n$-gluon amplitudes,
$S[\cdots ]$  is the momentum kernel 
(see \eqn{momentumkernel} below for the explicit definition)
and 
$\sigma$, $\tau$ range over all permutations of $\{4, \cdots, n\}$.
The field-theory formula (\ref{KLT}) was a harbinger 
of more recent developments showing that 
tree-level gravitational amplitudes
can be obtained as a double copy of gauge-theory 
amplitudes by replacing the 
color factors $c_i$ that appear
in a cubic decomposition of gauge-theory amplitudes
with kinematic numerators $\tilde{n}_i$ 
that obey the same algebraic relations
\cite{Bern:2008qj,Bern:2010yg}.
The classical spacetime background itself 
can be constructed through a double-copy 
procedure \cite{Monteiro:2014cda,Luna:2015paa,Luna:2016due,Ridgway:2015fdl,
Luna:2016hge}.
Much evidence has accumulated \cite{
Bern:2010ue,Carrasco:2011mn,Bern:2011rj,Bern:2012uf,Yuan:2012rg,Boels:2012ew,Carrasco:2012ca,Bjerrum-Bohr:2013iza,Bern:2013yya,Bern:2013uka,Nohle:2013bfa,Bern:2014sna}
for the conjecture \cite{Bern:2008qj,Bern:2010ue}
that color-kinematic duality and the double-copy procedure
also apply to the integrands of loop-level amplitudes,
but difficulties remain \cite{Bern:2017yxu,Bern:2017ucb}.
Quite recently, the KLT formula itself
has been generalized to one-loop gravitational 
amplitudes \cite{He:2016mzd,He:2017spx}.
\para

In 1999, Bern, De Freitas, and Wong \cite{Bern:1999bx}
proposed an expression analogous to \eqn{KLT} for 
the color-encoded tree-level $n$-gluon amplitude
\be
\Agluon
~= ~-~ \sum_{\sigma, \tau \in S_{n-3}} 
A (1,3, \sigma, 2)  ~S[\sigma|\tau]_3 ~\A (2,  3,   \tau, 1)
\label{dualKLT}
\ee
in terms of partial gauge-theory amplitudes
$A(\cdots)$ 
and dual partial scalar amplitudes
$\A(\cdots)$,
which are obtained from partial gauge-theory  amplitudes
by replacing kinematic numerators $n_i$
with color factors $c_i$ \cite{Bern:2010yg}.
Subsequently proven in ref.~\cite{Du:2011js} 
using BCFW recursion relations \cite{Britto:2005fq},
\eqn{dualKLT} makes manifest that 
a subset of $(n-3)!$ of the partial amplitudes 
is sufficient to produce the full color-encoded amplitude.\footnote{The relations (\ref{dualKLT}) can be viewed as a generalization of the
factorization observed in refs.~\cite{Zhu:1980sz,Goebel:1980es,Brown:1982xx}.}
Consequently, the full set of partial amplitudes
$A(\cdots)$ can be expressed in terms of 
these $(n-3)!$ independent partial amplitudes $A(1,3,\sigma,2)$;
these are the  well-known Bern-Carrasco-Johansson (BCJ) 
relations \cite{Bern:2008qj}.
 \para

In addition to being gauge invariant,
tree-level gauge-theory amplitudes have been shown to 
possess a color-factor symmetry \cite{Brown:2016mrh,Brown:2016hck}.
For each external gluon in the amplitude,
there is a family of momentum-dependent shifts of the
color factors $c_i$ that leave the amplitude invariant.
These shifts are analogous to generalized gauge transformations
of the kinematic numerators $n_i$
but more restrictive because they preserve
the Jacobi identities satisfied by color factors
(whereas generalized gauge transformations 
can relate Jacobi-satisfying kinematic numerators
to non-Jacobi-satisfying kinematic numerators).
The dual partial amplitudes $\A(\cdots)$
appearing in \eqn{dualKLT},
which depend on the color factors $c_i$,
are themselves invariant under color-factor shifts.
Thus \eqn{dualKLT} represents a decomposition
of the amplitude in terms of building blocks 
that are simultaneously gauge invariant
and color-factor symmetric.
\para

In this paper we use the color-factor symmetry 
of gauge-theory amplitudes to provide 
an alternative derivation of the KLT-type formula
(\ref{dualKLT}) for the $n$-gluon amplitude. 
As a by-product, we obtain a new KLT-type formula for the 
biadjoint scalar theory \cite{Cachazo:2013iea}
\be
\Abiad
~= ~-~ \sum_{\sigma, \tau \in S_{n-3}} 
\tA (1,3, \sigma, 2)  ~S[\sigma|\tau]_3 ~\A (2,  3,   \tau, 1)
\label{biadjointKLT}
\ee
a theory which also possesses color-factor symmetry \cite{Brown:2016mrh}.
\para

We also obtain new KLT-type relations
for tree-level $n$-point QCD amplitudes $\Ank$ containing 
$k$ differently flavored quark-antiquark pairs and $n-2k$ gluons.
Tree-level QCD amplitudes can be expressed in terms of 
partial amplitudes that obey group-theory
relations \cite{Kleiss:1988ne,Melia:2013bta,Melia:2013epa}
as well as (for amplitudes containing gluons) BCJ 
relations \cite{Johansson:2015oia}.
Johansson and Ochirov (JO) used these relations to define an 
independent basis of partial amplitudes $A(1, \gamma, 2)$,
where $\gamma$ denotes 
a particular subset (described in the main body of this paper)
of permutations of the remaining labels\footnote{Here 
$\{2, 4, \cdots, 2k\}$
denote the labels of (differently flavored) quarks,
$\{1, 3,  \cdots,  2k-1$\}
the labels of the corresponding antiquarks,
and $\{2k+1, \cdots, n\}$ the labels of gluons.}
of quarks and gluons $\{3, \cdots, n\}$.
For two or fewer quark-antiquark pairs,
the number of independent amplitudes is $(n-3)!$,
and the JO basis is simply given by $A (1, 3, \sigma, 2)$,
where $\sigma$  is any permutation of $\{4, \cdots, n\}$.
We establish that, in this case,
the color-encoded QCD amplitude can be expressed as 
\be
\Ankl 
~=~-~ \sum_{\sigma, \tau \in S_{n-3}} 
A (1,3, \sigma, 2)  ~S[\sigma|\tau]_3 ~\A (2,  3,   \tau, 1)
\ee
where 
$ A (1,3, \sigma, 2)$
are QCD partial amplitudes,
$\A (2,3, \tau, 1)$ are corresponding dual partial amplitudes,
and 
$ S[\sigma|\tau]_3 $
is the same momentum kernel that appears in the
all-gluon expression (\ref{dualKLT}),
modified by masses in the case of $k=2$ amplitudes.
For $k>2$, 
the number of independent amplitudes in the JO basis is 
$(n-3)! (2k-2)/k!$ \cite{Johansson:2015oia},
and the QCD amplitude can be expressed as 
\be
\Ank
~=~ \sum_{\gamma, \delta \in \JO} 
A (1, \gamma,  2)  
~T(1 \gamma 2| 2 \delta 1)
~ \A (2, \delta, 1)
\ee
where $\gamma$ and $\delta$ both belong to the JO set of permutations,
and $T( \cdots )$ is the 
inverse of a particular submatrix 
of double-partial amplitudes defined later in the paper.
Although we do not present an explicit expression for 
$T(\cdots)$,
we conjecture that it can be expressed as an 
$(n-3)^{\rm th}$ degree polynomial of kinematic invariants,
similar to the momentum kernel.
We also write analogous expressions for gravitational scattering amplitudes.
Earlier work on extensions of KLT relations to more general
gravitational amplitudes includes 
refs.~\cite{Bern:1999bx,Feng:2010br,Damgaard:2012fb,delaCruz:2016wbr}.
\para

This paper is structured as follows:
in sec.~\ref{sec:bicolor}, 
we introduce the bicolor scalar theory,
containing both biadjoint and bifundamental fields.
We describe the Melia basis of partial amplitudes,
the Melia-Johansson-Ochirov decomposition of the bicolor amplitude,
and finally the double-partial amplitudes of the bicolor theory.
We end with a KLT-type relation for amplitudes containing only
bifundamental fields.
In sec.~\ref{sec:cfs},
we show that the color-factor symmetry possessed by the bicolor theory
can be used to derive the null vectors of the matrix of 
double-partial amplitudes, which leads to BCJ relations for
the bicolor partial amplitudes.
In sec.~\ref{sec:kltbicolor}, we use the color-factor symmetry
to obtain a KLT-type relation for arbitrary bicolor amplitudes. 
In sec.~\ref{sec:kltqcd}, we derive KLT-type relations
for QCD amplitudes.
We also discuss more general gravitational KLT relations.
Section \ref{sec:concl} contains our conclusions.

\section{Bicolor scalar theory}
\setcounter{equation}{0}
\label{sec:bicolor}

The biadjoint scalar theory,
introduced by Cachazo, He, and Yuan in ref.~\cite{Cachazo:2013iea},
is a theory consisting of massless scalar particles
$ \phi_{\textsf{a}\textsf{a'}}$
transforming in the adjoint representation of the color group 
$U(N) \times U(\tilde{N})$
with cubic interactions of the form
\be
f^{\textsf{abc}}\tilde f^{\textsf{a'b'c'}}
\phi_{\textsf{a}\textsf{a'}}\phi_{\textsf{b}\textsf{b'}}\phi_{\textsf{c}\textsf{c'}}
\label{triplephi}
\ee
where
$f^{\textsf{abc}}$ and $\tilde f^{\textsf{a'b'c'}}$
are the structure constants of $U(N)$ and $U(\tilde{N})$.
Whereas gravity is a double copy of gauge theory
(replacing color factors $c_i$ with kinematic numerators $\tilde{n}_i$),
the biadjoint theory can be viewed as a zeroth copy of gauge theory
(replacing kinematic numerators $n_i$ with color factors  $\tilde{c}_i$). 
The double-partial amplitudes of the biadjoint theory,
which depend only on kinematic invariants of the external momenta
(without the complications of spin),
provide the cleanest examples of amplitudes obeying
Kleiss-Kuijf \cite{Kleiss:1988ne} and BCJ relations.
The biadjoint theory is also color-factor symmetric \cite{Brown:2016mrh},
which we will use in sec.~\ref{sec:kltbicolor} 
to derive KLT-type relations (\ref{biadjointKLT})
for its amplitudes.
\para

Because our goal is also to obtain KLT-type relations
for QCD amplitudes containing quarks as well as gluons,
we generalize the biadjoint scalar theory to include, 
in addition to the massless biadjoint fields, 
several flavors\footnote{This generalization 
was considered earlier in ref.~\cite{Naculich:2014naa}
for a single flavor of bifundamental scalar.
See also ref.~\cite{Anastasiou:2016csv}.}
of (possibly massive) scalar fields
$\psi_{(\s)}^{\textsf{i}\,\textsf{i'}}$,
$ \s=1, \cdots, N_f$,
transforming in the $R \otimes \tilde{R}$ representation of
$U(N) \times U(\tilde{N})$
with mass terms
\be
m_{(\s)}^2
~\overline\psi_{(\s)\textsf{i}\,\textsf{i'}}
\psi_{(\s)}^{\textsf{i}\,\textsf{i'}} 
\ee
as well as cubic couplings
\be
(T^{\textsf{a}})^{\textsf{i}}_{~\textsf{j}}
(\tilde T^{\textsf{a'}})^{\textsf{i'}}_{~\textsf{j'}}
\overline\psi_{(\s) \textsf{i}\,\textsf{i'}} \phi_{\textsf{a}\textsf{a'}}  
\psi_{(\s)}^{\textsf{j}\,\textsf{j'}} 
\label{psiphipsi}
\ee
where $(T^{\textsf{a}})^{\textsf{i}}_{~\textsf{j}}$
and $ (\tilde T^{\textsf{a'}})^{\textsf{i'}}_{~\textsf{j'}}$
are generators in the $R$ and $\tilde{R}$ representations.
For convenience in what follows, we will refer to 
$\psi_{(\s)}^{\textsf{j}\,\textsf{j'}} $
as bifundamental fields
(and $\overline\psi_{(\s) \textsf{i}\,\textsf{i'}}$ 
as anti-bifundamental fields), 
although the representation could be more general.
We refer to this as the bicolor scalar theory.
\para

\begin{figure}[t]
\begin{center}
\includegraphics[scale=1.0,trim=80 710 50 65,clip=true]{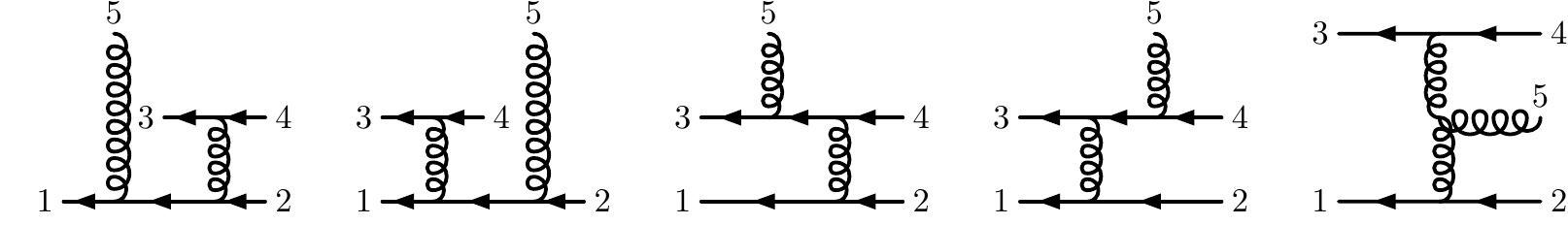}
\caption{\small Cubic diagrams $i=1$ through 5 for $\Abi_{5,2}$.
Lines with arrows denote bifundamental scalars
whereas curly lines denote biadjoint scalars.}
\label{fig:u2}
\end{center}
\end{figure}

Consider a tree-level $n$-point amplitude 
with both bifundamental and biadjoint fields
\be
\Abink (
\barpsi_1, \psi_2, \barpsi_3, \psi_4, \cdots, \barpsi_{2k-1}, \psi_{2k},
\phi_{2k+1}, \cdots, \phi_n)
\ee
where external fields 
$\psi$ in the bifundamental representation have even labels 
and fields $\barpsi$ in the anti-bifundamental representation 
have odd labels.
We assume that the $\psi_{2\ell}$ all have different flavors 
(and possibly different masses), 
with $\barpsi_{2\ell-1}$ having the corresponding antiflavor 
(and equal mass) to $\psi_{2\ell}$.
This amplitude is given by a sum over cubic diagrams
\be
\Abink~=~ \sum_{i \in {\rm cubic}}{c_i ~ \tc_i \over d_i} 
\label{abi}
\ee
where $c_i$, $\tc_i$ are color factors
constructed from the cubic vertices
(\ref{triplephi}) and (\ref{psiphipsi}),
and $d_i$ is the product of massless $\phi$ and massive $\psi$ propagators.
The cubic diagrams appearing in \eqn{abi} 
correspond to a subset of the cubic diagrams appearing in 
an $n$-point amplitude  of biadjoint fields.
For example, the five-point amplitude with 
two pairs of bifundamentals is given by
\be
\Abi_{5,2} ( \barpsi_1, \psi_2, \barpsi_3, \psi_4, \phi_5)
~=~ \sum_{i=1}^5 {c_i ~ \tc_i \over d_i } 
\label{fivepointcubicdecomp}
\ee
where the five contributing cubic diagrams are shown in fig.~\ref{fig:u2},
and the color factors and denominators have the 
form \cite{Johansson:2015oia,Brown:2016mrh}
\begin{align}
  c_1 &= (T^{\ta_5} T^\tb )^{\ti_1}_{~\ti_2} (T^\tb)^{\ti_3}_{~\ti_4} \,, \qquad~
& d_1 &= (s_{15}\!-\!m_1^2) s_{34} = 2 s_{34} \,  k_1 \cdot k_5 \, \,, \nn\\
  c_2 &= (T^\tb T^{\ta_5})^{\ti_1}_{~\ti_2} (T^\tb)^{\ti_3}_{~\ti_4} \,, \qquad~
& d_2 &= (s_{25}\!-\!m_1^2) s_{34} = 2 s_{34} \,  k_2 \cdot k_5 \, \,, \nn\\
  c_3 &= (T^\tb )^{\ti_1}_{~\ti_2} (T^{\ta_5} T^\tb)^{\ti_3}_{~\ti_4} \,, \qquad~
& d_3 &= s_{12} (s_{35}\!-\!m_3^2) = 2 s_{12} \, k_3 \cdot k_5 \, \,, \nn\\
  c_4 &= (T^\tb )^{\ti_1}_{~\ti_2} (T^\tb T^{\ta_5})^{\ti_3}_{~\ti_4} \,, \qquad~
& d_4 &= s_{12} (s_{45}\!-\!m_3^2) = 2 s_{12} \, k_4 \cdot k_5 \, \,, \nn\\
  c_5 &= f^{\ta_5 \tb c}\;\!  (T^\tb)^{\ti_1}_{~\ti_2} (T^c)^{\ti_3}_{~\ti_4} \,, \qquad~
& d_5 &= s_{12} s_{34} 
\label{fivepointcolorfactors}
\end{align} 
where $s_{ij}= (k_i + k_j)^2$,
with analogous expressions for $\tc_i$.
The color factors obey the Jacobi identities
\be
c_1 - c_2 + c_5 = 0, \qquad\qquad  c_3 - c_4 - c_5 = 0\,.
\ee
This five-point amplitude will be our prototypical example throughout the paper
as it nicely illustrates many of the features of bicolor amplitudes.
\para

Following ref.~\cite{Cachazo:2013iea}, 
we define partial amplitudes 
with respect to each color group factor 
as well as double-partial amplitudes.
The partial amplitude
$\tA (\alpha)$
with respect to the first group factor $U(N)$, 
where $\alpha$ denotes an arbitrary permutation
of the external particle labels $\{1, \cdots, n\}$, 
receives contributions 
from those cubic diagrams $i$ whose color factor $c_i$ 
can be drawn in a planar fashion with the external legs
in the cyclic order specified by the permutation $\alpha$. 
For example, 
by examining fig.~\ref{fig:u2},
one can write down  the following five-point partial 
amplitudes\footnote{For the remainder of the paper, 
we omit the commas between arguments for conciseness.}
\be
  \tA(15342)
 ~=~ \frac{\tc_1}{d_1} - \frac{\tc_3}{d_3} - \frac{\tc_5}{d_5} \,, 
\qquad
  \tA(13542)
 ~=~ \frac{\tc_3}{d_3} + \frac{\tc_4}{d_4} \,, 
\qquad 
   \tA(13452)
~=~ \frac{\tc_2}{d_2} - \frac{\tc_4}{d_4} + \frac{\tc_5}{d_5} 
\label{fivetwomelia}
\ee
where the $\pm$ sign in front of each term 
results from the antisymmetry of the structure constants 
$f^{\textsf{abc}}= -f^{\textsf{bac}}$
and a similar antisymmetry imposed on the generators
$(T^{\textsf{a}})^{\textsf{i}}_{~\textsf{j}}=
-(T^{\textsf{a}})_{\textsf{j}}^{~\textsf{i}} $
(see ref.~\cite{Johansson:2015oia}).
In general the partial amplitudes are given by 
\be
\tA (\alpha) ~=~ \sum_i  {M_{i,\alpha} \tc_i \over d_i}
\label{firstpartial}
\ee
where $M_{i,\alpha}$ vanishes if $c_i$ does not contribute to 
$\tA(\alpha)$,
and is otherwise given by 1 or $-1$. 
These partial amplitudes may be regarded as ``dual'' to 
color-ordered gauge-theory amplitudes,
as they can be obtained from the latter by 
replacing the kinematic numerators $n_i$ with $\tc_i$ \cite{Bern:2010yg}.
\para

When $k \ge 2$,
some of the partial amplitudes vanish 
because none of the cubic diagrams can contribute. 
In the five-point amplitude above, for example,
$\tA( 13245 ) $ vanishes because any 
contributing diagram would require the lines for 
the differently flavored bifundamental fields to cross,
and are thus non-planar. 
Moreover, there are group-theoretic relations among 
the nonvanishing bicolor partial amplitudes 
analogous to the Kleiss-Kuijf \cite{Kleiss:1988ne}
and Melia \cite{Melia:2013bta,Melia:2013epa}
relations among gauge-theory partial amplitudes.
For the five-point amplitude, some of the Melia relations are
\begin{align}
\tA(15432) &= - \tA(15342) - \tA(13542) \,,
\nn\\
\tA(14532) 
&= \tA(13452) \,,
\nn\\
\tA(14352) 
&=
- \tA(13452) - \tA(13542) \,.
\end{align}
These group-theoretic relations can be used to define 
an independent basis\footnote{The Melia basis amplitudes
are independent only with respect to purely group-theoretic relations.  
As we will see in sec.~\ref{sec:cfs}, 
color-factor symmetry implies additional (BCJ) relations among 
these amplitudes.}
of $(n-2)!/k!$ partial amplitudes,
called the Melia basis \cite{Melia:2013bta,Melia:2013epa}.
\para

To describe the Melia basis of partial amplitudes,
we recall that 
a Dyck word of length $2r$ is a string composed of  
$r$ letters $\X$ and $r$ letters $\Y$
such that the number of $\X$'s preceding any point in the string
is greater than the number of preceding $\Y$'s.
An easy way to understand this is to 
visualize $\X$ as a left bracket $\{$ 
and $\Y$ as a right bracket $\}$, 
in which case a Dyck word corresponds to a well-formed set of brackets.
The number of such words is $(2r)!/(r+1)!r!$, the $r^{\rm th}$ Catalan number. 
For example 
for $r=1$ there is only one Dyck word: $\{ \} $, 
for $r=2$ there are two: $\{\} \{\} $ and $\{\{\} \} $, 
and for $r=3$ there are five:
$\{\} \{\} \{\} $, 
$\{\} \{\{\} \} $, 
$\{\{\} \} \{\} $,
$\{\{\} \{\} \} $, and
$\{\{\{\} \} \} $. 
Consider the set of partial amplitudes 
$ A( 1, \gamma(3), \cdots, \gamma(n), 2 )$,
where $\gamma$ is any permutation of 
$\{ 3, \cdots, n  \}$ 
such that the set of $k-1$ $\X$ and $k-1$ $\Y$ 
in $\gamma$ form a Dyck word of length $2k-2$.
The biadjoint fields may be distributed anywhere 
among the $\barpsi$ and $\psi$ in $\gamma$. 
The  number of distinct allowed patterns of 
$\barpsi$, $\psi$, and $\phi$ is given by the number of Dyck words
of length $2k-2$ times the number of ways of distributing
$n-2k$ biadjoint fields among the letters of the Dyck word
\be 
{(2k-2)! \over k!(k-1)! }
\times 
{n-2 \choose  2k-2}  \,.
\ee
For each allowed pattern, 
there are $(n-2k)!$ distinct 
choices for the biadjoint labels,
and $(k-1)!$ choices for the $\barpsi$ labels.
The label on each $\psi$ is then fixed: 
it must have the flavor of the nearest unpaired $\barpsi$ to its left.
Thus, for example,
for $\Abi_{6,3}$ the allowed permutations $\gamma$ are
$ \barpsi_3 \psi_4 \barpsi_5 \psi_6 $,
$ \barpsi_5 \psi_6 \barpsi_3 \psi_4 $,
$ \barpsi_3 \barpsi_5 \psi_6 \psi_4 $, and
$ \barpsi_5 \barpsi_3 \psi_4 \psi_6 $,
whereas for $\Abi_{5,2}$ the allowed permutations are
$ \phi_5\barpsi_3 \psi_4 $,
$ \barpsi_3 \phi_5 \psi_4 $, and 
$ \barpsi_3 \psi_4 \phi_5$.
Thus the three partial amplitudes given in \eqn{fivetwomelia}
precisely comprise the Melia basis for $\Abi_{5,2}$.
The multiplicity of the Melia basis is given by 
\be
{(2k-2)! \over k!(k-1)! }
\times 
{n-2 \choose  2k-2} 
\times 
(n-2k)! 
\times (k-1)! 
~=~ {(n-2)!  \over k!}
\ee
as found in ref.~\cite{Melia:2013epa}.
We refer to the allowed permutations of $\gamma$ as the {\it Melia set},
and the partial amplitudes $A(1\gamma 2)$ as the {\it Melia basis}.
For $k=0$ and $k=1$, 
the elements of $\gamma$ are all biadjoint scalars,
the Melia set consists of {\it all} permutations of 
$\{ 3, \cdots, n\}$,
and the Melia basis coincides with the Kleiss-Kuijf 
basis \cite{Kleiss:1988ne}.
\para

Since,
for $\gamma$ belonging to the Melia set,
the 
$\tA (1\gamma 2) $
form an independent basis of partial amplitudes
with respect to the first group factor $U(N)$,
the bicolor amplitude can be written in a 
proper decomposition \cite{Melia:2015ika}
\be
\Abink 
= \sum_{\gamma \in \MB} \tA (1\gamma 2)  ~\CF
\label{proper}
\ee
for some set of color factors $\CF$.
\Eqn{proper} follows from \eqns{abi}{firstpartial},
provided that these color factors satisfy
\be
c_i~=~ \sum_{\gamma \in \MB}  M_{i,1\gamma 2}  ~\CF \,.
\label{CFdef}
\ee
For purely biadjoint amplitudes, 
the $\CF$ are simply half-ladder color factors
\be
\CF ~\equiv~  \sum_{\tb_1,\ldots,\tb_{n{-}3}}
f^{\ta_1 \ta_{\gamma(3)} \tb_1}
f^{\tb_1 \ta_{\gamma(4)} \tb_2}
\cdots 
f^{\tb_{n{-}3} \ta_{\gamma(n)} \ta_2} \,, 
\qquad\qquad \gamma \in S_{n-2} \,,
\qquad\qquad k=0
\label{halfladder}
\ee
and \eqn{proper} is the 
Del Duca-Dixon-Maltoni decomposition \cite{DelDuca:1999rs}.
For amplitudes containing one pair of bifundamentals, 
the $\CF$ are also half-ladder color factors
along a bifundamental backbone \cite{Kosower:1987ic,Mangano:1988kk}
\be
\CF ~=~ \left( {T}^{\ta_{\gamma(3)}}{T}^{\ta_{\gamma(4)}}
\cdots {T}^{\ta_{\gamma(n)}} \right)^{\ti_1}_{~~ \ti_2} \,,
\qquad\qquad \gamma \in S_{n-2} \,,
\qquad\qquad k=1 \,.
\ee
For $k \ge 2$, 
the requisite color factors were constructed 
by Johansson and Ochirov \cite{Johansson:2015oia},
and so we refer to \eqn{proper} as the Melia-Johansson-Ochirov (MJO)
decomposition.\footnote{Recently, a one-loop version of the 
MJO decomposition has been developed in ref.~\cite{Kalin:2017oqr}.}
We do not give here the explicit expressions for the JO color factors 
(which may be found in 
refs.~\cite{Johansson:2015oia,Melia:2015ika,Brown:2016hck})
but for our five-point example $\Abi_{5,2}$, they reduce to 
\be
   C_{15342} ~=~ c_1 \,, \qquad
   C_{13542} ~=~ c_2 + c_4 \,, \qquad
   C_{13452} ~=~ c_2 
\label{fivepointJO}
\ee
where $c_i$ are defined in \eqn{fivepointcolorfactors}.
It is straightforward to verify that 
the MJO decomposition
\be
\Abi_{5,2}  ~=~ 
\tA(15342)\, C_{15342}  ~+~\tA(13542)\, C_{13542}  ~+~\tA(13452) \, C_{13452} 
\ee
agrees with \eqn{fivepointcubicdecomp}.
The general proof was given in ref.~\cite{Melia:2015ika}.
\para

Similarly, we define partial amplitudes $\A (\beta ) $ 
with respect to the second group factor $U(\tilde{N})$, 
\be
\A (\beta)~=~ \sum_i  {M_{i,\beta} c_i \over d_i}
\label{secondpartial}
\ee
corresponding to the sum over cubic diagrams $i$ whose color factor $\tc_i$ 
can be drawn in a planar fashion with the external legs
in the cyclic order specified by the permutation $\beta$. 
Finally, 
we define double-partial amplitudes
corresponding to a sum over diagrams that satisfy
both of these criteria simultaneously
\be
m(\alpha|\beta)
= \sum_i  {M_{i,\alpha} M_{i,\beta}  \over d_i} \,.
\label{doublepartial}
\ee
In our five-point example, the double-partial amplitudes for 
which both $\alpha$ and $\beta$ belong to the Melia basis
are given by 
\be
\begin{pmatrix}
m(15342|15342) &     
m(15342|13542) &     
m(15342|13452) 
\\[2mm]
m(13542|15342) &     
m(13542|13542) &     
m(13542|13452) 
\\[2mm]
m(13452|15342) &     
m(13452|13542) &     
m(13452|13452) 
\end{pmatrix}
=
\begin{pmatrix}
{1\over d_1} +{1\over d_3} +{1\over d_5} &     -{1 \over d_3}   &  -{1 \over d_5} \\[2mm]
            -{1 \over d_3}  & {1\over d_3} +{1\over d_4} &  - {1 \over d_4} \\[2mm]
-{1 \over d_5}  &  - {1 \over d_4} & {1\over d_2} +{1\over d_4} +{1\over d_5}  \\[2mm]
\end{pmatrix} \,.
\label{fivepointdoublepartial}
\ee
For purely biadjoint amplitudes \cite{Cachazo:2013iea}
and for bicolor amplitudes with a single pair of 
bifundamentals \cite{Naculich:2014naa}, 
the double-partial amplitudes are equal to 
the elements of the propagator matrix defined in ref.~\cite{Vaman:2010ez}.
For bicolor amplitudes containing two or more (massless)
$\psi \bar{\psi}$ pairs,
the matrix of double-partial amplitudes will be 
a ``thinned-out'' version of the propagator matrix, 
because some of the cubic diagrams that contribute to 
the purely biadjoint amplitude will
be absent.
\para

Using \eqns{CFdef}{doublepartial}
we may express the partial amplitudes (\ref{secondpartial}) as
\be
\A (\beta) 
= \sum_{\gamma \in \MB} m(\beta |1 \gamma 2) ~\CF
\label{secondpartialredux}
\ee
and in particular the partial amplitudes 
belonging to the Melia basis are 
\be
\A (1 \delta 2) 
= \sum_{\gamma \in \MB} m(1\delta 2|1 \gamma 2) ~\CF \,,
\qquad\qquad \delta \in \MB \,.
\ee
For amplitudes containing only bifundamental fields ($n=2k$),
the $(2k-2)!/k! ~\times~ (2k-2)!/k!$ matrix whose entries
are given by 
$m(1\delta 2|1 \gamma 2) $
may be inverted to give
\be
\CF~=~ \sum_{\delta \in \MB}
m^{-1} (1\gamma 2|1 \delta 2) ~\A (1 \delta  2) , 
\qquad\qquad \hbox{for $n=2k$} \,.
\label{invertm}
\ee
This may in turn be inserted into \eqn{proper}  to yield
\begin{align}
\Abi_{2k,k}
&= 
\sum_{\gamma, \delta \in \MB} 
\tA (1 \gamma 2) ~m^{-1} (1\gamma 2|1 \delta 2) ~\A (1 \delta  2) 
\label{matinverse}
\end{align}
which has the structure of a KLT-type relation 
for the bicolor amplitude.
The elements of the inverse matrix 
$m^{-1} (1\gamma 2|1 \delta 2) $
are rational functions of the kinematic invariants,
but we conjecture that,
using momentum conservation,
they can be written as polynomials in the kinematic invariants.
(We have verified this for $\Abi_{4,2}$ and $\Abi_{6,3}$.)
\para

When the amplitude contains biadjoint fields
as well as bifundamentals ($n>2k$),
the rank of the matrix $m(1\delta 2|1 \gamma 2) $
is less than $(n-2)!/k!$,
and therefore the matrix cannot be inverted.
It possesses null vectors: eigenvectors with eigenvalue zero.
In the next section, we will use the color-factor symmetry
to determine these null vectors.
Then in sec.~\ref{sec:kltbicolor} 
we will identify an invertible submatrix of this matrix,
and thereby write a KLT-type relation for 
bicolor amplitudes that contain both bifundamental and biadjoint fields.
\para

\section{Color-factor symmetry of the bicolor scalar theory} 
\setcounter{equation}{0}
\label{sec:cfs} 

A new symmetry of gauge-theory amplitudes was
introduced in ref.~\cite{Brown:2016mrh}, 
one which acts on the color factors $c_i$  
while leaving the amplitude invariant.
Specifically, color factors undergo 
momentum-dependent shifts $\delta c_i$ 
that preserve the Jacobi identities satisfied by the color factors.
The proof of invariance of gauge-theory amplitudes
under color-factor shifts employed a decomposition
called the radiation vertex expansion.
\para

Ref.~\cite{Brown:2016hck} 
examined the color-factor symmetry of QCD amplitudes
involving $k$ massive quark-antiquark pairs and $n-2k$ gluons.
For each of the external gluon legs $a$ in a QCD amplitude,
there is a family of color-factor shifts $\dela c_i$.
It was shown that the color factors $\CF$ defined by Johansson and Ochirov
transform in a natural way under color-factor shifts
\be
\dela~  C_{1 \sigma(3) \cdots \sigma(b-1) a \sigma(b) \cdots  \sigma(n) 2 }
~=~
 \alpha_{a,\sigma} 
\left( k_a \cdot k_1  +  \sum_{c=3}^{b-1} k_a \cdot k_{\sigma(c)} \right)
\label{delaC}
\ee
where $\alpha_{a,\sigma}$ is a set of arbitrary, independent
parameters (or functions) associated with the family of shifts,
with $\sigma$ denoting a fixed permutation 
of the remaining legs $ \{3, \cdots, n \} \setminus  \{ a\}$
that belongs to the Melia basis.
To give a specific example, 
the five-point amplitude $\cA^{\rm qcd}_{5,2}$ 
is invariant under a one-parameter shift $\delta_{5} c_i$ 
under which the JO color factors transform as
\begin{align}
\delta_{5} C_{15342} &= \alpha_{5,34} ~k_5 \cdot k_1\,, \nn \\
\delta_{5} C_{13542} &= \alpha_{5,34} ~k_5 \cdot  (k_1+k_3) \,, 
\label{fivepointshifts}
\\
\delta_{5} C_{13452} &= \alpha_{5,34}~ k_5 \cdot (k_1+k_3+k_4) \,. \nn
\end{align}
In general, since the number of gluons in the amplitude is 
$n-2k$ and the number of Melia permutations $\sigma$ 
is $(n-3)!/k!$ 
the number of independent color-factor shifts is given by
$(n-2k) (n-3)!/k!$ for $k \ge 2$.
(For $k=0$ and $k=1$, the number of independent 
shifts is $(n-3) (n-3)!$)
\para

In sec.~8 of ref.~\cite{Brown:2016mrh} it was shown that amplitudes of
the biadjoint scalar theory are also invariant under color-factor shifts,
which was proven using the cubic vertex expansion.
This proof is straightforwardly extended to the amplitudes of 
the bicolor scalar theory introduced in the previous section.
\para

The color-factor symmetry of bicolor amplitudes can be used to 
derive the null eigenvectors of the matrix of 
double-partial amplitudes.
First we use  \eqn{CFdef} and its analog for $\tc_i$ to 
express the bicolor amplitude (\ref{abi}) 
in terms of the JO color factors 
\be
\Abink= 
\sum_{\gamma, \delta \in \MB} 
\tCF
~m(1\delta 2|1 \gamma 2) 
~\CF\,.
\ee
Its variation under a shift of the color factors $\dela c_i$ is  
\be
\dela \Abink ~=~ 
\sum_{\gamma, \delta \in \MB} 
\tCF
~m(1\delta 2|1 \gamma 2) 
~\dela \CF \,.
\ee
Since the amplitude is invariant under color-factor shifts,
and since $\tCF$ constitute an independent basis, 
we conclude that $\dela \CF$ are null
vectors of the matrix $ m(1\delta 2|1 \gamma 2) $
\be
0~=~ 
\sum_{\gamma \in \MB} 
m(1\delta 2|1 \gamma 2) 
~\dela \CF \,.
\label{invarianceofpartial}
\ee
Given the independence of $\alpha_{a,\sigma}$ 
we obtain
\be
\sum_{b=3}^{n+1} \left( k_a \cdot k_1 + \sum_{c=3}^{b-1} k_a \cdot k_{\sigma(c)}  \right)
~m(1 \delta 2| 1 \sigma(3) \cdots \sigma(b-1) a \sigma(b) \cdots \sigma(n) 2 )
~=~ 0
\label{nullvectors}
\ee
i.e., we have derived a set of null eigenvectors of the  matrix $m(1\delta 2|1 \gamma 2)$.
Since the matrix of double-partial amplitudes is symmetric
(cf. \eqn{doublepartial}),
we can also write
\be
\sum_{b=3}^{n+1} \left( k_a \cdot k_1 + \sum_{c=3}^{b-1} k_a \cdot k_{\sigma(c)}  \right)
~m(1 \sigma(3) \cdots \sigma(b-1) a \sigma(b) \cdots \sigma(n) 2 |1 \gamma 2)
~=~ 0 \,.
\label{nullvectorsredux}
\ee
The number of independent null eigenvectors is simply 
the number of independent color-factor shifts specified above, 
namely $(n-2k) (n-3)!/k!$ for $k \ge 2$,
and $(n-3) (n-3)!$ for $k=0$ and $k=1$.
Subtracting this from the size of the Melia basis $(n-2)!/k!$,
we obtain the rank of the matrix of double-partial amplitudes
\cite{Johansson:2015oia}
\be
\beta(n,k)~=~
\begin{cases}
(n-3)! &  k = 0, 1, 2, \\
(n-3)! (2k-2)/k!  \qquad &  k \ge 2\,.
\end{cases}
\label{beta}
\ee
The existence of null eigenvectors 
has two consequences for the partial amplitudes
\be
\A (\beta) 
= \sum_{\gamma \in \MB} m(\beta |1 \gamma 2) ~\CF \,.
\label{secondpartialrepeat}
\ee
First,
by virtue of \eqn{invarianceofpartial}
they are invariant under the color-factor symmetry
\be
\dela \A (\beta) 
= \sum_{\gamma \in \MB} m(\beta |1 \gamma 2) ~\dela \CF =0 \,.
\label{partialinvariant}
\ee
Second, 
by virtue of \eqn{nullvectorsredux},
the set of partial amplitudes $\A(1 \gamma 2)$
(where $\gamma$ belongs to the Melia set) 
are not independent, as they obey the fundamental BCJ relations
\be
\sum_{b=3}^{n+1} \left( k_a \cdot k_1 + \sum_{c=3}^{b-1} k_a \cdot k_{\sigma(c)}  \right)
~\A (1, \sigma(3), \cdots, \sigma(b-1), a, \sigma(b), \cdots, \sigma(n), 2 )
~=~ 0 \,.
\label{qcdbcj}
\ee
These relations reduce the number of independent amplitudes to $\beta(n,k)$.
\para

Using the fundamental BCJ relations (\ref{qcdbcj}),
Johansson and Ochirov identified an independent basis of
$\beta(n,k)$ amplitudes for $k\ge 2$, 
namely those $\A(1\gamma 2)$
for which $\gamma$ belongs to the Melia set
{\it and} with $\gamma(3)$ restricted to be one of the $\bar{\psi}$ fields, 
i.e. $\gamma(3)$ belongs to the set $\{3, 5, 7, \cdots 2k-1\}$.
We refer to this subset of permutations $\gamma$ as the {\it JO set}.
In the 
$\Abi_{5,2} ( \barpsi_1, \psi_2, \barpsi_3, \psi_4, \phi_5)$
example,
of the three Melia partial amplitudes
$\A(15342)$, $\A(13542)$, and $\A(13452)$,
only the last two belong to the JO basis.
The JO set of permutations will play a prominent role in the remainder
of this paper, in which we use the color-factor symmetry to express
color-encoded amplitudes in terms of the JO basis.
\para

For $k=2$ amplitudes, 
$\gamma(3)$ can only be 3, 
so the JO basis consists of the $(n-3)!$ amplitudes 
$A(1,3,\sigma,2)$, 
in which $\sigma$ is an arbitrary permutation of $\{4, \cdots, n\}$.
The same set of $(n-3)!$ amplitudes
provides an independent  basis for $k=0$ and $k=1$ amplitudes as well,
in which $3$ labels one of the biadjoint fields.
\para

\section{KLT-type relations for biadjoint and bicolor amplitudes}
\setcounter{equation}{0}
\label{sec:kltbicolor}

Having set the stage by introducing the bicolor scalar theory 
and its amplitudes in sec.~\ref{sec:bicolor}, 
and by exploring the consequences of their invariance under 
color-factor symmetry in sec.~\ref{sec:cfs}, 
we now derive a KLT-type relation for biadjoint and bicolor amplitudes.
The results of this section will facilitate the derivation of 
KLT-type relations for QCD amplitudes 
in sec.~\ref{sec:kltqcd}. 
\para

We first perform a color-factor shift 
\be
C'_{1\gamma 2}~=~ C_{1\gamma 2} + \delta C_{1\gamma2}
\label{JOshift}
\ee
to set to zero {\it all} 
color factors not belonging to 
the JO set of permutations $\gamma$.
(Color-factor redefinitions of this form were 
previously  discussed in sec.~5 of ref.~\cite{BjerrumBohr:2010hn}
for all-gluon amplitudes.)
For example, for $\Abi_{5,2}$ we choose
$\alpha_{5,34} = - C_{15342}/k_5 \cdot k_1$ in 
\eqn{fivepointshifts}
to obtain
\begin{align}
C'_{15342} &= 0 \,,\nn \\
C'_{13542} &= 
C_{13542} - { k_5 \cdot  (k_1+k_3) \over k_5 \cdot k_1 } C_{15342} \,,
\\
C'_{13452} &= 
C_{13452} -  {k_5 \cdot (k_1+k_3+k_4)\over k_5 \cdot k_1} C_{15342}   \,.
\nn
\end{align}
In the appendix, we compute the analogous color-factor shift
for the five-point biadjoint amplitude $\Abi_{5,0}$
(or the  five-gluon amplitude $\cA^{\rm qcd}_{5,0}$)
from which the procedure for the general $n$-point amplitude
will be clear.
Since the bicolor amplitude (\ref{proper}) 
is invariant under the color-factor shift (\ref{JOshift}), 
we may write it as 
\be
\Abink 
= \sum_{\gamma \in \JO} \tA (1\gamma 2)  ~\CFp 
\label{bicolorJO}
\ee
where the sum is now restricted to the JO set of permutations.
\para

An alternative way to obtain the shifted color factors $\CFp$
employs the fact that $\tA (1\gamma 2) $ with $\gamma \in \JO$
is an independent basis, 
and that BCJ relations can be used to write all the 
other partial amplitudes $\tA( 1\gamma 2)$ with $\gamma \in \MB$
in terms of the JO basis,
thus reducing \eqn{proper} to the form of \eqn{bicolorJO} for
some appropriate set of $C'_{1\gamma 2}$.
Johansson and Ochirov use this approach to obtain 
\eqn{bicolorJO} with 
an explicit expression for $C'_{1\gamma2}$
(cf. eq.~(4.45) of ref.~\cite{Johansson:2015oia}).
\para

Since the partial amplitudes (\ref{secondpartialredux})
are invariant under the color-factor shift  
(cf. \eqn{partialinvariant}), 
we can also write them as a sum over the JO set of permutations
\be
\A (\beta )
= \sum_{\gamma \in \JO} m(\beta |1 \gamma 2) ~\CFp \,.
\label{secondpartialJO}
\ee
Let us now restrict our attention to amplitudes with two or fewer
pairs of bifundamental fields. 
For $k \le 2$, 
permutations belonging to the JO set must have $\gamma(3)=3$,
and so \eqn{secondpartialJO}  takes the form 
\be
\A (\beta )
= \sum_{\sigma \in S_{n-3}} m(\beta |1 3 \sigma 2) ~\CFps \,,
\quad\quad k \le 2.
\ee
(For $k=0$ and 1, this is true by definition.
For $k=2$, this is because 1 and 3 label the only $\bar{\psi}$
fields in the amplitude.)
We further restrict our attention to the particular set
of partial amplitudes
\be
\A (2 3 \tau 1)
= \sum_{\sigma \in S_{n-3}} m(23\tau 1 |1 3 \sigma 2) ~\CFps \, ,
\quad\quad k \le 2
\label{restrictedpartial}
\ee
where $\tau$ is an arbitrary permutation
of $\{ 4, \cdots, n\}$.
\para

\begin{figure}[t]
\begin{center}
\includegraphics[scale=0.3]{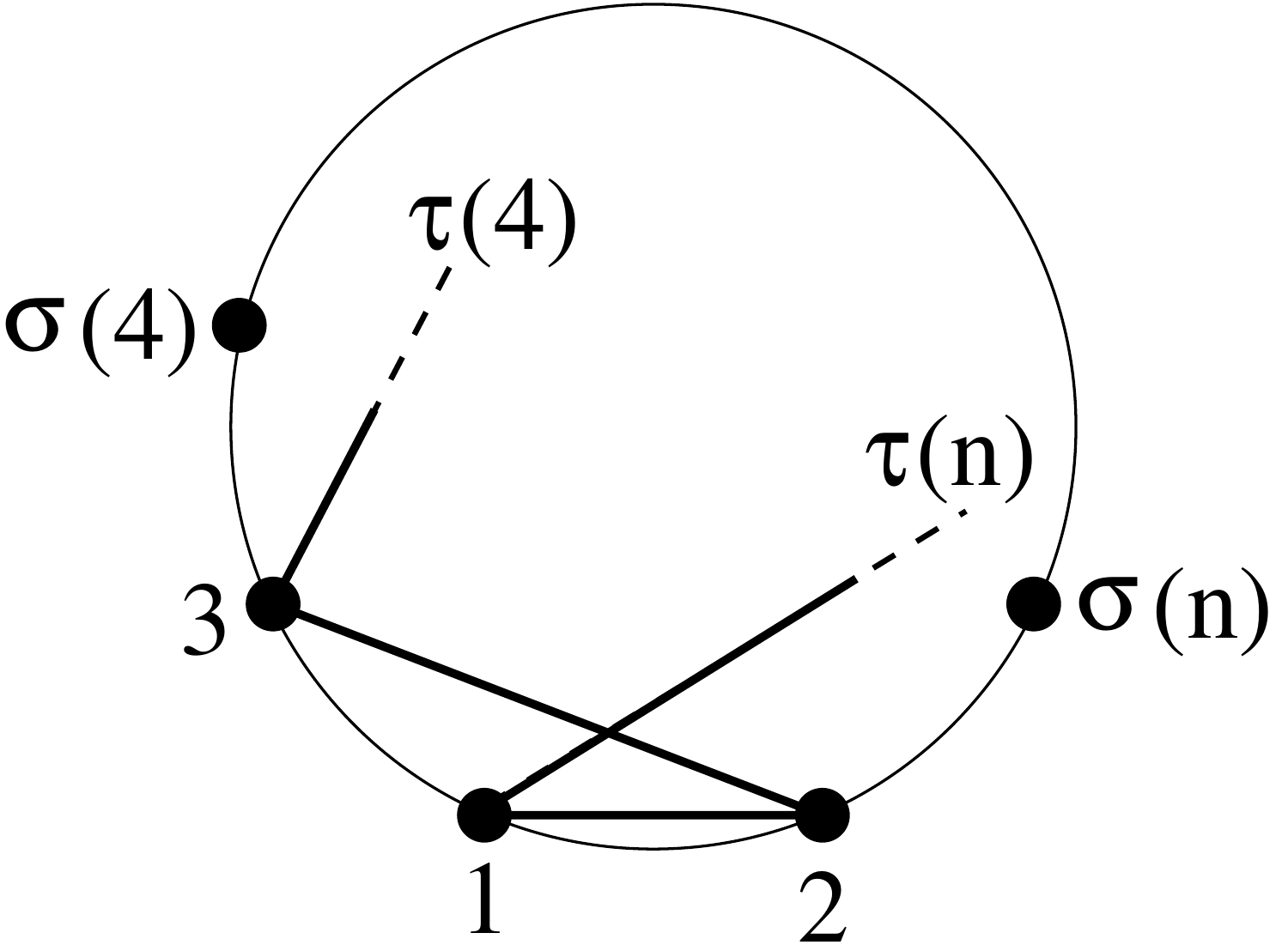}
\caption{\small CHY prescription for computing the double-partial amplitude
$m(2 3 \tau  1 | 1 3 \sigma  2)$.}
\label{fig:chy}
\end{center}
\end{figure}

Let us now examine the $(n-3)! \times (n-3)!$ 
matrix of double-partial amplitudes
$m(2 3 \tau  1 | 1 3 \sigma  2)$ 
for amplitudes with $k \le 2$, beginning with $k=0$.
For the biadjoint scalar theory,
Cachazo, He, and Yuan gave an algorithm \cite{Cachazo:2013iea} 
for computing double-partial amplitudes 
that begins with drawing a circle with the labels 
$\{1$, $3$, $\sigma(4), \cdots, \sigma(n)$, $2 \}$
around the perimeter,
and then inscribing a polygon whose vertices are 
$\{2$, $3$, $\tau(4), \cdots, \tau(n)$,$ 1\}$  inside,
as shown in fig.~\ref{fig:chy}.
Because the segments of the polygon emerging from 1 and 2 
must cross, the CHY prescription dictates that 
$m(2 3 \tau  1 | 1 3 \sigma  2)$ 
is given by $1/s_{12}$
times an $(n-1)$-point  double-partial amplitude
$m(v 3 \tau   | v 3 \sigma )$,
in which one of the fields ($v$) is off-shell.
Since $k_v$ can be eliminated 
in terms of $k_3, \cdots, k_n$
using momentum conservation,
the double-partial amplitudes $m(v 3 \tau   | v 3 \sigma )$ 
can be expressed in terms of invariants $s_{ab}$ with $3 \le a, b \le n$.
For example, for the amplitude $\Abi_{5,0}$, one has
\be
\begin{pmatrix}
m(23541|13542) &     
m(23541|13452) 
\\[2mm]
m(23451|13542) &     
m(23451|13452) 
\end{pmatrix}
= 
{1 \over  s_{12}} \begin{pmatrix}
 - {1 \over s_{35} } - {1 \over  s_{45}} & 
   {1 \over  s_{45}} \\[2mm]
   {1 \over  s_{45}} &
 - {1 \over  s_{45}} - {1 \over  s_{34}}
\end{pmatrix} \,.
\label{fivepointprop}
\ee
The inverse of this matrix (using momentum conservation) is given by
\be
\begin{pmatrix}
-s_{35} (s_{34}+s_{45})  
&
- s_{34} s_{35}
\\[2mm]
- s_{34} s_{35} 
&
-s_{34} (s_{35}+s_{45}) 
\end{pmatrix}
~=~
~-~\begin{pmatrix}
S[54|54]_3
&
S[54|45]_3
\\[2mm]
S[45|54]_3
&
S[45|45]_3
\end{pmatrix}
\ee
where $S[\sigma|\tau]_3$ is the momentum
kernel \cite{BjerrumBohr:2010ta,BjerrumBohr:2010zb,BjerrumBohr:2010yc,BjerrumBohr:2010hn} defined as\footnote{
Here $\sigma, \tau\in S_{n-3}$ are permutations acting on 
labels $\{4,\cdots,n\}$. 
Define $\theta(r,s)_\tau=1$ if the ordering of $r,s$ 
is the same in both sequences of labels, $\{\sigma(4),\cdots,\sigma(n)\}$ 
and $\{\tau(4), \cdots, \tau(n)\}$, and zero otherwise.
The original 
definition \cite{BjerrumBohr:2010ta,BjerrumBohr:2010zb,BjerrumBohr:2010yc,BjerrumBohr:2010hn} 
of the momentum kernel is slightly 
modified \cite{Broedel:2013tta,Cachazo:2013iea}
to be symmetric in its arguments,
$S[\sigma|\tau]_3 = S[\tau|\sigma]_3$.
}
\be
S[\sigma|\tau]_3 ~=~
\prod^{n}_{i=4}\left[s_{3, \sigma(i)}+\sum^{i{-}1}_{j=4} \theta(\sigma(j), \sigma(i))_{\tau} ~s_{\sigma(j),\sigma(i)}\right] \,.
\label{momentumkernel}
\ee
For $n$-point amplitudes in general, the inverse of 
$m(2 3 \tau  1 | 1 3 \sigma  2)$ 
is given by the negative of the momentum kernel,
$- S[\sigma|\tau]_3$,
as was shown by Cachazo, He, and Yuan \cite{Cachazo:2013iea} 
by using KLT orthogonality \cite{Cachazo:2013gna}.
This can be seen from the results of ref.~\cite{Cachazo:2013iea} 
by relabeling the external legs
$1 \to 3$, $n-1 \to 1$, $n \to 2$, and $\sigma(i) \to \sigma(i+2)$, 
and then using cyclic symmetry of the double-partial
amplitudes 
$m(3 \tau 1 2 | 3 \sigma 2 1) = m(2 3 \tau  1 | 1 3 \sigma  2)$.
\para

Next consider bicolor amplitudes with one pair of 
(possibly massive) bifundamentals, $\barpsi_1$ and $\psi_2$.
It was shown in ref.~\cite{Naculich:2014naa}
that the double-partial amplitudes with $k=1$, 
when expressed in terms of 
$k_a \cdot k_b$ with $2 \le a, b \le n$, 
are identical to those of the biadjoint theory.
Above we showed that the particular double-partial amplitudes 
$m(2 3 \tau  1 | 1 3 \sigma  2)$
can be written\footnote{using $s_{12} = (k_3 + \cdots k_n)^2$}
in terms of
$k_a \cdot k_b$ with $3 \le a, b \le n$. 
Since $s_{ab}= 2k_a \cdot k_b$ for $3 \le a, b \le n$, 
the matrix of double-partial amplitudes
$m(2 3 \tau  1 | 1 3 \sigma  2)$
with $k=1$ is identical,
when written in terms of 
$s_{ab}$ with $3 \le a, b \le n$,
to that of the biadjoint scalar theory,
and therefore has the same inverse, namely $-S[\sigma|\tau]_3$.
\para

Next let us consider bicolor amplitudes with two pairs of bifundamentals, 
$\barpsi_1$, $\psi_2$ and $\barpsi_3$, $\psi_4$,
with masses $m_1=m_2$ and $m_3=m_4$.
In general, fewer cubic diagrams will contribute to 
these double-partial amplitudes
relative to purely biadjoint double-partial amplitudes
because some of the cubic vertices present in the latter
are ruled out by flavor conservation, etc. 
(\eg, $\barpsi_1 \barpsi_3 \phi_a$ and $\barpsi_1 \psi_4 \phi_a$).
For the specific subclass of double-partial amplitudes 
$m(2 3 \tau  1 | 1 3 \sigma  2)$, however, 
the same set of cubic diagrams that contribute in the
purely biadjoint case will also contribute to $k=2$ 
bicolor amplitudes.
This is because
for these amplitudes,
as we explained above, 
the fields $\barpsi_1$, $\psi_2$ 
are effectively replaced
by a virtual biadjoint field $\phi_v$,
and the $n$-point double-partial amplitude  
is given by $1/s_{12}$ times an $(n-1)$-point 
double-partial amplitude 
with $(n-3)$ external biadjoint fields, one virtual biadjoint field,
and one bifundamental pair $\barpsi_3$, $\psi_4$.
These latter (effectively $k=1$) 
double-partial amplitudes are equal to
the analogous biadjoint double-partial amplitudes, 
except that we must replace $s_{ij}$ with  $s'_{ij}$,
where
\begin{align}
s'_{12} &= s_{12} ,&
s'_{34} &= s_{34} ,&
s'_{a b} &= s_{a b},
\nn\\
s'_{3a} &= s_{3a} - m_3^2,&
s'_{4a} &= s_{4a} - m_3^2,&
   a, b  & \ge 5.
\label{sprimedef}
\end{align}
For example, for $\Abi_{5,2}$, \eqn{fivepointprop} becomes
\be
\begin{pmatrix}
m(23541|13542) &     
m(23541|13452) 
\\[2mm]
m(23451|13542) &     
m(23451|13452) 
\end{pmatrix}
= 
{1 \over  s'_{12}} \begin{pmatrix}
 - {1 \over s_{35}'} - {1 \over  s'_{45}} & 
   {1 \over  s'_{45}} \\[2mm]
   {1 \over  s'_{45}} &
 - {1 \over  s'_{45}} - {1 \over  s'_{34}}
\end{pmatrix}
\ee
as can be verified by 
examining \eqns{fivepointcolorfactors}{fivepointdoublepartial}.
In general, for $k=2$ amplitudes, the 
inverse of $m(2 3 \tau  1 | 1 3 \sigma  2)$
is given by (minus) the momentum kernel (\ref{momentumkernel})
with $s_{ij} \to s'_{ij}$.
\para

Thus, for all bicolor amplitudes with $k\le 2$,
we have established that the inverse of the matrix 
of double-partial amplitudes 
$m(2 3 \tau  1 | 1 3 \sigma  2)$ 
is given by (minus) the momentum kernel,
$-S[\sigma|\tau]_3$
(with $s_{ij} \to s'_{ij}$ in the $k=2$ case).
Hence  \eqn{restrictedpartial} 
can be inverted to give an explicit expression for the shifted color factors
\be
\CFps ~=~
~-~ \sum_{\tau \in S_{n-3}}
S[\sigma|\tau]_3 
~ \A (2 3 \tau 1) \,,
\qquad\qquad k \le 2 \,.
\label{invertS}
\ee
In turn, \eqn{invertS} can be inserted into \eqn{bicolorJO} 
to obtain the following expression for $k\le 2$ bicolor amplitudes\footnote{
Recall that for $k=2$ we must let $s_{ij} \to s'_{ij}$ 
in \eqn{momentumkernel}.} 
\be
\boxed{
\Abinkl
~=~ ~-~ \sum_{\sigma, \tau \in S_{n-3}} 
\tA (13\sigma 2)  ~S[\sigma|\tau]_3 ~\A (2 3  \tau 1)  
}
\,.
\ee
This is our new KLT-type relation 
for biadjoint and bicolor scalar amplitudes (with $k\le 2$),
which expresses the full amplitude in terms of 
dual partial scalar amplitudes $\A (\cdots)  $ and $\tA (\cdots)$.
For example, for the five-point amplitude we have
\be
\Abi_{5,2} ~=~
\begin{pmatrix}
\tA(13542),  & \tA(13452)
\end{pmatrix}
\begin{pmatrix}
-s'_{35} (s'_{34}+s'_{45})  
&
- s'_{34} s'_{35}
\\[2mm]
- s'_{34} s'_{35} 
&
-s'_{34} (s'_{35}+s'_{45}) 
\end{pmatrix}
\begin{pmatrix}
\A(23541) \\[2mm]
\A(23451) 
\end{pmatrix}\,.
\ee
We can also write an expression 
for the most general partial scalar amplitude $\A (\beta )$
in terms of the partial amplitudes belonging to the independent basis 
$\A (2 3  \tau 1)$,
\begin{align}
\A (\beta )
&~=~ ~-~ \sum_{\sigma, \tau \in S_{n-3}} 
m(\beta |1 3 \sigma 2) 
~S[\sigma|\tau]_3 ~\A (2 3  \tau 1) \,,
\qquad\qquad k \le 2
\end{align}
by inserting \eqn{invertS} into  \eqn{secondpartialJO}.
These are precisely BCJ relations for scalar partial amplitudes.
\para

For amplitudes with $k > 2$, 
the fundamental BCJ relations imply that the matrix of double-partial
amplitudes has rank $\beta(n,k)$, defined in \eqn{beta}.
In general, the $\beta(n,k) \times \beta(n,k)$ submatrix 
$m(2 \delta  1|1 \gamma 2) $,
where both  $\delta$ and $\gamma$ belong to the JO set,
should be invertible.
Provided this is the case, the equation\footnote{
The partial amplitudes
$\A (2 \delta 1 )$
constitute an alternative JO basis 
with opposite choice of ``signature'' 
\cite{Melia:2013epa}
for the $\barpsi_1$, $\psi_2$ bifundamental pair 
relative to the usual JO basis.}
\be
\A (2 \delta 1 )
~=~ \sum_{\gamma \in \JO} m(2 \delta 1 |1 \gamma 2) ~\CFp\,,
\qquad\qquad \delta \in \JO
\ee
can be inverted to give
\be
\CFp ~=~
\sum_{\delta \in \JO} 
T(1 \gamma 2| 2 \delta 1)
~ \A (2 \delta 1)\,,
\qquad\qquad \gamma \in \JO
\label{invertT}
\ee
although, unlike the $k \le 2$ case,
we cannot present at this point an explicit expression for the 
inverse matrix $T$.
Inserting \eqn{invertT} into \eqn{bicolorJO}, 
we can write a KLT-type relation for a general bicolor amplitude
\be
\Abink 
~=~ \sum_{\gamma, \delta \in \JO} 
\tA (1 \gamma 2)  
~T(1 \gamma 2| 2 \delta 1)
~ \A (2 \delta 1) \,.
\ee
Similarly, by inserting \eqn{invertT} into \eqn{secondpartialJO},
we obtain the BCJ relations 
\begin{align}
\A (\beta )
& ~=~ \sum_{\gamma, \delta \in \JO} 
m(\beta |1 \gamma  2) 
~T(1 \gamma 2| 2 \delta 1)
~ \A (2 \delta 1) \,.
\end{align}
We anticipate that, like the momentum kernel,
the matrix elements of 
$T(1 \gamma 2| 2 \delta 1)$ will be $(n-3)^{\rm th}$ order polynomials
in the kinematic invariants.

\section{KLT-type relations for QCD (and gravity) amplitudes} 
\setcounter{equation}{0}
\label{sec:kltqcd}

In this section,
we derive KLT-type relations for QCD amplitudes 
with quarks and gluons in two different ways.
The first way uses the invariance of QCD amplitudes
under color-factor symmetry \cite{Brown:2016mrh,Brown:2016hck},
together with the results of the previous section.
The second way explicitly invokes color-kinematic duality
\cite{Bern:2008qj,Johansson:2015oia}.
Both methods of course yield the same result.
In the final subsection, we apply our methods to
obtain KLT relations for gravitational amplitudes.
\para

\subsection{Derivation of QCD KLT-type relations using color-factor symmetry}

An $n$-point tree-level QCD amplitude with $k$ differently flavored
(massive) quarks $\psi$, 
$k$ (massive) anti-quarks $\barpsi$ with corresponding anti-flavors,
and $(n-2k)$ gluons
is given by a sum over cubic diagrams
\be
\Ank (\barpsi_1, \psi_2, \barpsi_3, \psi_4, \cdots, \barpsi_{2k-1}, \psi_{2k},
g_{2k+1}, \cdots, g_n)
~=~ 
\sum_{i \in {\rm cubic}}{c_i ~ n_i \over d_i} 
\label{ank}
\ee
where $c_i$ and $d_i$ are the same color factors
and propagators appearing in sec.~\ref{sec:bicolor}
and $n_i$ are kinematic numerators for QCD.
Using \eqn{CFdef}, 
we can rewrite \eqn{ank} in the 
Melia-Johansson-Ochirov proper 
decomposition \cite{
Melia:2013bta,Melia:2013epa,Johansson:2015oia,Melia:2015ika}
\be
\Ank
~=~ \sum_{\gamma \in \MB} A (1\gamma 2) ~\CF 
\label{MJO}
\ee
where
$\CF$ are the JO color factors described in sec.~\ref{sec:bicolor}
and 
\be
A(\alpha) ~=~ \sum_i  {M_{i,\alpha} n_i \over d_i}
\label{colorordered}
\ee
are color-ordered partial amplitudes of the gauge theory.
\para

For amplitudes containing one or more gluons, 
$\Ank$ is invariant under a family of color-factor shifts,
as was shown in ref.~\cite{Brown:2016mrh}
by expanding the amplitude in a radiation vertex expansion.
Invariance of \eqn{MJO} under a color-factor shift $\dela c_i$ 
implies 
\be
\dela \Ank ~=~ \sum_{\gamma \in \MB} A(1\gamma 2) ~\dela \CF 
~=~0
\label{delaA}
\ee
where $\dela \CF$ are given in  \eqn{delaC}.
Since the parameters $\alpha_{a,\sigma}$ in \eqn{delaC} 
are independent, \eqn{delaA} implies
\be
\sum_{b=3}^{n+1} \left( k_a \cdot k_1 + \sum_{c=3}^{b-1} k_a \cdot k_{\sigma(c)}  \right)
A (1, \sigma(3), \cdots, \sigma(b-1), a, \sigma(b), \cdots, \sigma(n), 2 )
~=~ 0 
\label{qcdBCJ}
\ee
which are simply the fundamental 
BCJ relations for color-ordered QCD amplitudes,
derived using color-factor symmetry.
These relations were first discovered 
for all-gluon amplitudes in ref.~\cite{Bern:2008qj},
and proven in 
refs.~\cite{BjerrumBohr:2009rd,Stieberger:2009hq,Feng:2010my,Chen:2011jxa}.
The fundamental BCJ relations were extended to QCD amplitudes 
for all values of $k$ in ref.~\cite{Johansson:2015oia},
and subsequently proven in ref.~\cite{delaCruz:2015dpa}.
They were shown to be a consequence of color-factor symmetry
in refs.~\cite{Brown:2016mrh,Brown:2016hck}.
\para

Using the color-factor symmetry  of the amplitude, 
we can rewrite \eqn{MJO} 
in terms of shifted color factors $\CFp$
defined in sec.~\ref{sec:kltbicolor},
which vanish unless $\gamma$ belongs to the JO set of permutations
\be
\Ank
~=~ \sum_{\gamma \in \JO } A (1\gamma 2) ~\CFp\,.
\ee
For amplitudes with $k\le 2$, 
we can use \eqn{invertS} to  
write\footnote{
For $k=2$, one must let $s_{ij} \to s'_{ij}$ 
in \eqn{momentumkernel}, 
where $s'_{ij}$ are defined in \eqn{sprimedef}.}
\be
\boxed{
\Ankl
~=~-~ \sum_{\sigma, \tau \in S_{n-3}} 
A (13\sigma 2)  ~S[\sigma|\tau]_3 ~\A (2 3  \tau 1)
}
\,.
\label{qcdKLT}
\ee
This is our KLT-type expression for QCD amplitudes
with two or fewer quark-antiquark pairs.
For $k=0$, 
an expression equivalent to this 
first appeared in 
ref.~\cite{Bern:1999bx} and was proven in 
ref.~\cite{Du:2011js} using BCFW techniques \cite{Britto:2005fq}.
In this paper, we have established that 
\eqn{qcdKLT} is also valid for amplitudes with $k=1$ and $k=2$.
We emphasize that each of the terms in the sum 
is both gauge-invariant (the partial gauge-theory amplitudes) 
and color-factor symmetric (the dual partial scalar amplitudes),
as is, of course, the entire color-encoded amplitude.
\para

For amplitudes with more than two quark-antiquark pairs,
we use \eqn{invertT} to obtain 
\be
\Ank
= \sum_{\gamma, \delta \in \JO} 
A (1 \gamma 2)  
~T(1 \gamma 2| 2 \delta 1)
~ \A (2 \delta 1)
\label{qcdKLTT}
\ee
where 
$T(1 \gamma 2| 2 \delta 1)$ 
is the inverse of the matrix
$m(2 \delta  1|1 \gamma 2) $,
for both $\delta$ and $\gamma$ belonging to the JO set.
For all-quark amplitudes ($n=2k$), 
there is no color-factor symmetry,
and \eqn{invertm} may be used
to rewrite \eqn{MJO} as 
\begin{align}
\cA^{\rm qcd}_{2k,k}
&= 
\sum_{\gamma, \delta \in \MB} 
A (1 \gamma 2) ~m^{-1} (1\gamma 2|1 \delta 2) ~\A (1 \delta  2)  \,.
\end{align}
We cannot present at this point explicit expressions for 
$T$ or $m^{-1}$ in these two equations, 
but we anticipate that they take the form of 
$(n-3)^{\rm th}$ order polynomials of kinematic invariants.
\para

\subsection{Derivation of QCD KLT-type relations using generalized gauge invariance} 

An alternative proof of the KLT-type relations for QCD amplitudes
begins by assuming color-kinematic duality,
which means that the kinematic numerators satisfy the same algebraic
relations as the color factors \cite{Bern:2008qj,Johansson:2015oia}.
The Jacobi relations among the color factors $c_i$
allow them to be written in terms of 
a set of independent color factors $\CF$ 
as in \eqn{CFdef}.
The analogous kinematic Jacobi equations allow the kinematic numerators $n_i$
to be written as
\be
n_i~=~ \sum_{\gamma \in \MB}  M_{i,1\gamma 2}  ~ \NF 
\label{Ndef}
\ee
for some set of independent numerators $\NF$.
Expressions for $\NF$ in terms of $n_i$ 
parallel those for $\CF$ in terms of $c_i$ 
\cite{Johansson:2015oia}.
We use \eqn{Ndef}, together with \eqn{secondpartial},
to write \eqn{ank} as
\be
\Ank = \sum_{\gamma \in \MB} \A (1\gamma 2) ~ \NF
\label{fullN}
\ee
as was done for all-gluon amplitudes ($k=0$) in ref.~\cite{Bern:2010yg}.
Similarly, the color-ordered partial amplitudes
(\ref{colorordered}) can be written as 
\be
A (\alpha) = \sum_{\gamma \in \MB} m(\alpha |1 \gamma 2) ~\NF
\label{partialN}
\ee
using \eqn{doublepartial}.
\para

Kinematic numerators $n_i$ that satisfy 
the same algebraic relations as color factors 
$c_i$ can necessarily be written in the form (\ref{Ndef}),
but this requirement does not uniquely determine the $\NF$
(unless there are no gluons, $n=2k$).
Generalized gauge transformations on $n_i$
(provided they preserve the Jacobi relations)
can result in a different set of $\NF$'s.
In ref.~\cite{Brown:2016mrh}, 
we described a set of {\it restricted} 
generalized gauge transformations 
of all-gluon amplitudes 
that preserve the Jacobi relations,
analogous to the shifts of color factors. 
In the context of QCD amplitudes of gluons and quarks,
for each of the external gluon legs $a$,
there is a family of 
restricted generalized gauge transformations
$\tdela  n_i$.
These restricted generalized gauge transformations act on
the independent numerators $\NF$ as
\be
\tdela~  N_{1 \sigma(3) \cdots \sigma(b-1) a \sigma(b) \cdots  \sigma(n) 2 }
~=~
 \beta_{a,\sigma} 
\left( k_a \cdot k_1  +  \sum_{c=3}^{b-1} k_a \cdot k_{\sigma(c)} \right)
\ee
where $\beta_{a,\sigma}$ is a set of arbitrary, independent
parameters (or functions),
with $\sigma$ denoting a fixed permutation 
of the remaining legs $ \{3, \cdots, n \} \setminus  \{ a\}$
that belongs to the Melia basis.
We can choose $\beta_{a,\sigma}$ in such a way that
the shifted numerators
\be
N'_{1\gamma 2}~=~ N_{1\gamma 2} + \tilde{\delta} N_{1\gamma2}
\label{shiftednumerators}
\ee
vanish except for those
in which $\gamma$ is restricted to the JO set.   
We described how to do this for color factors in sec.~\ref{sec:kltbicolor}
and in the appendix, and the procedure is the same in this case.
Since gauge-theory amplitudes are invariant under generalized gauge
transformations, 
\eqns{fullN}{partialN} can be written as 
\begin{align}
\Ank
&
= \sum_{\gamma \in \JO} \A (1\gamma 2) ~ \NFp\,,
\label{AJO}
\\
A (\alpha) 
&
= \sum_{\gamma \in \JO} m(\alpha |1 \gamma 2) ~\NFp
\label{colororderedJO}
\end{align}
where the sums are now restricted to the JO set of permutations.
\para

For amplitudes with two or fewer quark-antiquark pairs,
we may invert \eqn{colororderedJO}, just as we did 
in sec.~\ref{sec:kltbicolor}, 
to obtain 
\be
\NFps~=~
~-~ \sum_{\tau \in S_{n-3}}
S[\sigma|\tau]_3 
~ A (2 3 \tau 1) \,,
\qquad\qquad k \le 2 \,.
\label{invertN}
\ee
For all-gluon amplitudes ($k=0$),
these numerators are essentially those written down by
Kiermaier in ref.~\cite{Kiermaier:2010}, and later by Cachazo, He, 
and Yuan in ref.~\cite{Cachazo:2013iea}.
\para

By inserting \eqn{invertN} into \eqn{AJO},
we obtain the KLT-type relation 
\be
\Ankl
= ~-~ \sum_{\sigma, \tau \in S_{n-3}} 
\A (13\sigma 2)  ~S[\sigma|\tau]_3 ~A (2 3  \tau 1) \,,
\ee
which is equivalent to \eqn{qcdKLT}.
Similarly, for $k>2$, we can obtain an expression equivalent to
\eqn{qcdKLTT}.
\para

By inserting \eqn{invertN} into \eqn{colororderedJO},
we obtain an expression for the most general color-ordered
amplitude $A (\alpha )$
in terms of partial amplitudes belonging to the independent basis 
$A (2 3  \tau 1)$, namely
\begin{align}
A (\alpha )
&~=~ ~-~ \sum_{\sigma, \tau \in S_{n-3}} 
m(\alpha |1 3 \sigma 2) 
~S[\sigma|\tau]_3 ~A (2 3  \tau 1) \,,
\qquad\qquad k \le 2 \,.
\end{align}
In the all-gluon case ($k=0$), this is essentially the expression given 
in appendix C of ref.~\cite{Cachazo:2014xea}.
It is equivalent in content (for $k\le 2$)  to
eq.~(4.37) of ref.~\cite{Johansson:2015oia},
though different in form.

\subsection{Gravitational KLT amplitudes}

Having invoked in the previous subsection
color-kinematic duality for the QCD kinematic numerators $n_i$,
we can now use the double-copy prescription 
\cite{Bern:2008qj,Bern:2010ue,Bern:2010yg}
to obtain gravitational scattering amplitudes containing 
$n-2k$ gravitons and $2k$ matter 
particles \cite{Johansson:2014zca,Chiodaroli:2014xia,Johansson:2015oia}
\be
\Agvnk ~=~ 
\sum_{i \in {\rm cubic}}{n_i \tilde{n}_i \over d_i}  \,.
\ee
Color-kinematic duality then allows us to use \eqn{Ndef} to write
\be
\Agvnk
~=~ \sum_{\gamma \in \MB} \tilde{A} (1\gamma 2) ~\NF 
\label{gravN}
\ee
where $\tAYM (1\gamma 2)$ are 
QCD partial amplitudes (\ref{colorordered})
with $n_i$ replaced by $\tilde{n}_i$.
This expression was first introduced for all-graviton amplitudes ($k=0$)
in ref.~\cite{Bern:2010yg}.
By virtue of the fact that the QCD partial amplitudes $\tAYM (1 \gamma 2 )$ obey
BCJ relations (\ref{qcdBCJ}), 
the amplitude (\ref{gravN}) is invariant under (restricted)
generalized gauge transformations
\be
{\tilde \delta} \Agvnk
~=~ \sum_{\gamma \in \MB} \tAYM  (1\gamma 2) ~ {\tilde \delta} \NF  ~=~0 \,.
\ee
Hence we can write \eqn{gravN} in terms of shifted numerators (\ref{shiftednumerators})
\be
\Agvnk
~=~ \sum_{\gamma \in \JO} \tAYM  (1\gamma 2) ~\NFp
\ee
with the sum now restricted to the JO set of permutations.
We may then use \eqn{invertN} we obtain the gravitational KLT 
relation\footnote{Earlier work on extensions of KLT relations to more general
gravitational amplitudes includes refs.~\cite{Bern:1999bx,Feng:2010br,Damgaard:2012fb,delaCruz:2016wbr}.}
\be
\Agvnkl
= ~-~ \sum_{\sigma, \tau \in S_{n-3}} 
\tAYM  (13\sigma 2)  ~S[\sigma|\tau]_3 ~A (2 3  \tau 1) 
\ee
valid\footnote{Recall that for $k=2$ we must let $s_{ij} \to s'_{ij}$  
in \eqn{momentumkernel}.} 
for amplitudes with $k\le 2$. 
A similar expression can be written for amplitudes with $k>2$.
For $k=0$, this is just the original (field-theory) KLT relation
for the tree-level $n$-graviton amplitude,
so we have come full circle to the starting point of this paper.
\para

\section{Conclusions}
\setcounter{equation}{0}
\label{sec:concl}

Gauge-theory amplitudes possess a color-factor symmetry,
which acts on its color factors $c_i$  
via momentum-dependent shifts
while leaving the tree-level amplitudes 
invariant. 
A direct consequence of this symmetry 
are the fundamental BCJ relations satisfied by the 
color-ordered partial amplitudes of 
Yang-Mills theory 
and QCD. 
\para

The biadjoint scalar theory,  
which can be considered the zeroth copy of Yang-Mills theory,
also possesses color-factor symmetry. 
In this paper, we introduced another theory with color-factor symmetry,
the bicolor scalar theory.
This theory contains both massless biadjoint scalars 
as well as massive bifundamental scalars, 
and can be regarded as the zeroth copy of QCD.
The partial amplitudes of the biadjoint and bicolor 
scalar theories are dual to the partial amplitudes
of Yang-Mills and QCD, as they can be obtained by
replacing the kinematic numerators $n_i$ in the latter
with color factors $\tilde{c}_i$.
We showed that the dual partial amplitudes are themselves 
invariant under color-factor symmetry, 
and also obey BCJ relations.
\para

The color-factor symmetry was then used to recast 
tree-level biadjoint and bicolor amplitudes into a KLT-type form,
involving a sum over products of dual partial amplitudes 
multiplied by a momentum-dependent function $T(\cdots)$.
This momentum-dependent function
is given by the inverse of a particular submatrix
of double-partial amplitudes
of the bicolor theory,
and is thus a rational function of the kinematic invariants.
For amplitudes with two or fewer bifundamental pairs ($k\le 2$),
this function was shown to be an 
$(n-3)^{\rm th}$  degree polynomial
of the kinematic invariants,
specifically the momentum kernel of 
refs.~\cite{BjerrumBohr:2010ta,BjerrumBohr:2010zb,BjerrumBohr:2010yc,BjerrumBohr:2010hn,Broedel:2013tta,Cachazo:2013iea},
slightly modified by masses in the $k=2$ case.
We conjecture that, for amplitudes with $k>2$, 
the momentum-dependent function will also be an 
$(n-3)^{\rm th}$  degree polynomial
in the invariants, 
for which we hope that an explicit expression can be found.
\para

We also used the color-factor symmetry 
to obtain a new KLT-type relation for tree-level QCD amplitudes,
involving a sum over products of 
QCD partial amplitudes and dual partial amplitudes,
and the same momentum-dependent function
that appeared in the bicolor amplitude.
Each term in this sum is both gauge-invariant
and color-factor symmetric, as is the full color-ordered amplitude.
The KLT-type relation was then alternatively obtained 
through a derivation that explicitly 
invoked color-kinematic duality
and a restricted generalized gauge transformation.
Finally, the double-copy prescription  
together with the same generalized gauge transformation
was used to obtain the KLT relation for gravitational
amplitudes.
\para

By utilizing the color-factor symmetry
and its parallels with generalized gauge transformations,
we have presented a unified treatment of the
derivation of KLT-type relations for the tree-level amplitudes 
\begin{align}
\Abink 
&= \sum_{\gamma, \delta \in \JO} 
\tA (1 \gamma 2)  
~T(1 \gamma 2| 2 \delta 1)
~\A (2 \delta 1) \,,
\nn\\
\Ank
&= \sum_{\gamma, \delta \in \JO} 
A (1 \gamma 2)  
~T(1 \gamma 2| 2 \delta 1)
~\A (2 \delta 1) \,,
\nn\\
\Agvnk
&= \sum_{\gamma, \delta \in \JO} 
\tAYM  (1\gamma 2)  
~T(1 \gamma 2| 2 \delta 1)
~A (2 \delta 1) 
\end{align}
of the bicolor scalar theory,
QCD, and gravity.
\para

\section*{Acknowledgments}
This material is based upon work 
supported by the National Science Foundation under 
Grants Nos.~PHY17-20202 and PFI:BIC 1318206.
RWB is also supported by funds
made available through a CWRU Institute Professorship Chair.

\vfil\break

\section*{Appendix}
\setcounter{equation}{0}
\label{sec:app}

In this appendix, we describe the specific  color-factor shift 
required to write the five-gluon amplitude 
(or five-point biadjoint amplitude) 
in terms of BCJ basis amplitudes.
In the process, we obtain the five-point BCJ relations \cite{Bern:2008qj}.
Our procedure readily generalizes to an arbitrary $n$-point amplitude.
\para

The Del Duca-Dixon-Maltoni decomposition \cite{DelDuca:1999rs}
of the five-gluon amplitude
(or alternatively the five-point amplitude of the biadjoint scalar theory)
is given by 
\begin{align}
\cA_{5,0} 
~=~  \sum_{ \gamma  \in S_3}  A(1\gamma 2) ~C_{1\gamma 2}
~=~ & A(13452) ~C_{13452} + A(13542) ~C_{13542} +A(14352)  ~C_{14352}  \nn\\
~+~  &A(15342)  ~C_{15342}  +A(14532)  ~C_{14532} +A(15432)  ~C_{15432}
\label{fivepointamplitude}
\end{align}
where $A(1 \gamma 2)$ 
constitute the Kleiss-Kuijf basis of partial amplitudes
of the corresponding theory.
There is a four-parameter family of color-factor shifts 
generated by gluons 4 and 5
\begin{align}
\delta C_{13452} &~=~ \alpha_{4,35} ~k_4 \cdot (k_1 + k_3)       ~+~ \alpha_{5,34} ~k_5 \cdot (k_1 + k_3 + k_4)  \,,\nn\\
\delta C_{13542} &~=~ \alpha_{4,35} ~k_4 \cdot (k_1 + k_3 + k_5) ~+~ \alpha_{5,34} ~k_5 \cdot (k_1 + k_3)  \,,\nn\\
\delta C_{14352} &~=~ \alpha_{4,35} ~k_4 \cdot k_1               ~+~ \alpha_{5,43} ~k_5 \cdot (k_1 + k_3 + k_4)  \,,\nn\\
\delta C_{15342} &~=~ \alpha_{4,53} ~k_4 \cdot (k_1 + k_3 + k_5) ~+~ \alpha_{5,34} ~k_5 \cdot  k_1   \,,\nn\\
\delta C_{14532} &~=~ \alpha_{4,53} ~k_4 \cdot k_1               ~+~ \alpha_{5,43} ~k_5 \cdot (k_1 + k_4)  \,,\nn\\
\delta C_{15432} &~=~ \alpha_{4,53} ~k_4 \cdot (k_1 + k_5)       ~+~ \alpha_{5,43} ~k_5 \cdot k_1   \,.
\label{cfs}
\end{align}
We now show explicitly how to define a set of shifted color factors
\be
C'_{1\gamma 2} = C_{1\gamma 2} + \delta C_{1\gamma 2}
\ee
that vanish except for $C'_{13452}$ and $C'_{13542}$.
First we consider the color factors 
in which the label 3 is to the right of both 4 and 5.
We set
$\delta C_{14532}= - C_{14532}$
and 
$\delta C_{15432} = -C_{15432}$ 
by requiring 
\be
\begin{pmatrix}
k_4 \cdot  k_1              & k_5 \cdot (k_1 + k_4)  \\
k_4 \cdot (k_1+k_5)         & k_5 \cdot k_1  
\end{pmatrix}
\begin{pmatrix}
\alpha_{4,53} \\
\alpha_{5,43} 
\end{pmatrix}
= 
- 
\begin{pmatrix}
C_{14532} \\
C_{15432}
\end{pmatrix}
\ee
which can be solved to give
\be
\begin{pmatrix}
\alpha_{4,53} \\
\alpha_{5,43} 
\end{pmatrix}
= 
{1 \over (k_4 \cdot k_5) ( k_2 \cdot k_3)} 
\begin{pmatrix}
k_5 \cdot  k_1              &-k_5 \cdot (k_1 + k_4)  \\
-k_4 \cdot (k_1+k_5)         & k_4 \cdot k_1  
\end{pmatrix}
\begin{pmatrix}
C_{14532} \\
C_{15432}
\end{pmatrix} \,.
\label{firstalpha}
\ee
Next we set $\delta C_{14352}= - C_{14352}$
and $\delta C_{15342} = -C_{15342}$ by choosing
\begin{align}
\alpha_{4,35}   &= ~-~  {C_{14352}   \over k_4 \cdot k_1}     -  {k_5 \cdot (k_1 + k_3 + k_4)  \over k_4 \cdot k_1} \alpha_{5,43} \,, 
\nn\\
\alpha_{5,34}   &= ~-~  {C_{15342}   \over k_5 \cdot k_1}     -  {k_4 \cdot (k_1 + k_3 + k_5) \over k_5 \cdot  k_1} \alpha_{4,53} \,.
\label{secondalpha} 
\end{align}
Finally we plug \eqns{firstalpha}{secondalpha} into the 
first two equations of \eqn{cfs} to obtain, 
after the use of momentum conservation and some algebra,
\begin{align}
C'_{13452} 
&=
C_{13452} 
- {k_4 \cdot (k_1 + k_3) \over k_4 \cdot k_1} C_{14352}  
+ {k_5 \cdot k_2 \over k_5 \cdot k_1}      C_{15342}    
\nn\\
&    
\hskip2cm - {k_2 \cdot k_5 \over k_2 \cdot k_3} 
    \left(   {k_4 \cdot k_3 \over k_4 \cdot k_1 }  \right) C_{14532} 
+ {k_2 \cdot k_5 \over k_2 \cdot k_3} 
    \left( { k_4 \cdot k_2  \over k_5 \cdot  k_1} - 1 \right) C_{15432} \,,
\nn\\
  C'_{13542} 
&= 
  C_{13542} 
+ {k_4 \cdot k_2  \over k_4 \cdot k_1} C_{14352}        
- {k_5 \cdot (k_1 + k_3)  \over k_5 \cdot k_1} C_{15342}   
\nn\\
&
\hskip2cm 
+ {k_4 \cdot k_2 \over k_2 \cdot k_3}
     \left(  {k_5 \cdot k_2  \over  ~k_4 \cdot k_1 }  -1  \right) C_{14532}
- {k_4 \cdot k_2 \over k_2 \cdot k_3}
	\left(  {k_5 \cdot  k_3  \over  k_5 \cdot  k_1~ } \right)  C_{15432}
\,,
\nn\\
  C'_{14352} 
&= 
0 \,,
\nn\\
  C'_{15342} 
&= 
0 \,,
\nn\\
  C'_{14532} 
&= 
0 \,,
\nn\\
  C'_{15432} 
&= 
0 \,.
\label{shiftedfivepoint}
\end{align}
Since the amplitude is invariant under this color-factor shift, 
we have 
\begin{align}
\cA_{5,0} &= 
A(13452) ~C'_{13452} +  A(13542) ~C'_{13542}  \,.
\label{shiftedfivepointamplitude}
\end{align}
Substituting \eqn{shiftedfivepoint}
into \eqn{shiftedfivepointamplitude}
and equating to \eqn{fivepointamplitude}
we obtain
\begin{align}
A(14352)  
&~=~
    - {k_4 \cdot (k_1 + k_3) \over k_4 \cdot k_1} A(13452)
    + {k_4 \cdot k_2  \over k_4 \cdot k_1} A(13542) 
\,,
\nn\\
A(15342)  
&~=~
	{k_5 \cdot k_2 \over k_5 \cdot k_1}  A(13452)     
	- {k_5 \cdot (k_1 + k_3)  \over k_5 \cdot k_1}      A(13542) 
\,,
\nn\\ 
A(14532) 
&~=~
	-{k_2 \cdot k_5 \over k_2 \cdot k_3} 
	\left(   {k_4 \cdot k_3 \over k_4 \cdot k_1 }  \right) A(13452) 
     +   {k_4 \cdot k_2 \over k_2 \cdot k_3}
       \left(  {k_5 \cdot k_2  \over  ~k_4 \cdot k_1 }  -1  \right) A(13542)
\,,
\nn\\
A(15432)
&~=~
{k_2 \cdot k_5 \over k_2 \cdot k_3} 
	\left( { k_4 \cdot k_2  \over k_5 \cdot  k_1} - 1 \right)   A(13452)    
       -{k_4 \cdot k_2 \over k_2 \cdot k_3}
     \left(  {k_5 \cdot  k_3  \over  k_5 \cdot  k_1~ } \right)  A(13542)
\end{align}
which are precisely the BCJ relations 
for five-gluon partial amplitudes \cite{Bern:2008qj}.
\para

An analogous procedure can be used for any $n$-point QCD amplitude
to define shifted color factors
that vanish except for those belonging to the JO basis.

\vfil\break

\end{document}